\documentclass{ws-mpla}
\usepackage[super]{cite}
\usepackage{graphicx}

\usepackage{euscript,epsfig,amsmath,amssymb}
\usepackage{amsfonts,latexsym}

\newcommand{\be}[1]{\begin{equation}\label{#1}}
\newcommand{\ee}{\end{equation}}
\newcommand{\ba}[1]{\begin{eqnarray}\label{#1}}
\newcommand{\ea}{\end{eqnarray}}
\newcommand{\rf}[1]{(\ref{#1})}
\newcommand{\nn}{\nonumber}
\newcommand{\mc}[1]{\mathcal{#1}}
\newcommand{\mb}[1]{\mathbf{#1}}
\newcommand{\half}{\frac{1}{2}}
\renewcommand{\theequation}{\arabic{section}.\arabic{equation}}

\begin{document}

\markboth{Branislav Vlahovic, Maxim Eingorn \& Cosmin Ilie} {Uniformity of CMB as a non-inflationary geometrical effect}

\catchline{}{}{}{}{}

\title{UNIFORMITY OF COSMIC MICROWAVE BACKGROUND\\ AS A NON-INFLATIONARY GEOMETRICAL EFFECT}

\author{\footnotesize BRANISLAV VLAHOVIC}

\address{North Carolina Central University, CREST and NASA Research Centers,\\ Fayetteville st. 1801, Durham, North Carolina 27707, U.S.A.\\
vlahovic@nccu.edu}

\author{\footnotesize MAXIM EINGORN}

\address{North Carolina Central University, CREST and NASA Research Centers,\\ Fayetteville st. 1801, Durham, North Carolina 27707, U.S.A.\\
maxim.eingorn@gmail.com}

\author{\footnotesize COSMIN ILIE}

\address{Department of Physics and Astronomy, University of North Carolina at Chapel Hill,\\
CB\#3255, Phillips Hall, Chapel Hill, NC 27599, U.S.A.\\
cilie@unc.edu}

\maketitle

\pub{Received (Day Month Year)}{Revised (Day Month Year)}

\begin{abstract}
The conventional $\Lambda$CDM cosmological model supplemented by the inflation concept describes the Universe very well. However, there are still a few concerns:
new Planck data impose constraints on the shape of the inflaton potential, which exclude a lot of inflationary models; dark matter is not detected directly, and
dark energy is not understood theoretically on a satisfactory level. In this brief sketch we investigate an alternative cosmological model with spherical spatial
geometry and an additional perfect fluid with the constant parameter $\omega=-1/3$ in the linear equation of state. It is demonstrated explicitly that in the
framework of such a model it is possible to satisfy the supernovae data at the same level of accuracy as within the $\Lambda$CDM model and at the same time
suppose that the observed cosmic microwave background (CMB) radiation originates from a very limited space region. This is ensured by introducing an additional
condition of light propagation between the antipodal points during the age of the Universe. Consequently, the CMB uniformity can be explained without the
inflation scenario.
The corresponding drawbacks of the model with respect to its comparison with the CMB data are also discussed.

\end{abstract}

\ccode{PACS Nos.: 04.20.-q; 98.80.-k}

\section{Introduction}
Even if it was a very successful theory in its own right, predicting for instance the primordial abundance of light elements, Big Bang (BB) cosmology suffers from
serious theoretical drawbacks such as the horizon, flatness, and the grand unified theory (GUT) magnetic monopoles problems. Cosmic
inflation~\cite{Guth:1980,Linde:1981,Albrecht:1982} can solve all three of those shortcomings, and, moreover, provides a natural mechanism for causally generating
the observed nearly scale invariant spectrum of primordial adiabatic density fluctuations\footnote{Alternatives are the
ekpyrotic~\cite{Khoury:2001a,Khoury:2001b,Khoury:2001c} and cyclic models~\cite{Steinhardt:Cyclic,Steinhardt:CyclicEvol,Khoury:CyclicDesign}.}. It is a theory
that has gained enormous popularity since being introduced in the early 1980s. However, as we shall discuss in some detail in Sec.\ref{SS:ProblemsInfl}, inflation
faces some theoretical challenges that motivate the search for alternative solutions to the problems of BB cosmology.

Moreover there is the puzzle of the initial singularity and the question it raises immediately: what happened prior to the Big Bang? Historically, the first model
of cosmic inflation was discovered by Alexei Starobinsky in 1980, when he suggested the possibility of a non-singular cosmological model~\cite{Starobinsky:1980}
as a consequence of including one-loop quantum vacuum polarization effects due to conformally covariant matter fields on a Friedmann-Lemaitre-Robertson-Walker
(FLRW) background. In his model the early universe goes through a maximally symmetric, de Sitter phase, corresponding to an exponential growth of the scale
factor. This accelerated nearly exponential expansion is typical in any inflationary theory ever proposed. Mukhanov \& Chibisov~\cite{Mukhanov:1981} have
considered the quantum fluctuations of the metric during this intermediate de Sitter phase and shown that the spectrum is nearly scale invariant, with an
amplitude consistent with generating the observed large scale structure of the Universe. All subsequent successful models of inflation replicate those two
predictions.

In the remainder of this section we will give a brief review of inflationary cosmology, structured as follows: in Sec.~\ref{SS:PredictionsInfl} we expand on the
discussion of the mechanisms and generic predictions of inflationary cosmology; in Sec.~\ref{SS:TestsInfl} we argue how current Cosmic Microwave Background (CMB)
data can be used as a testbed for inflationary model building. We will end this introduction in Sec.~\ref{SS:ProblemsInfl} with a discussion of the theoretical
drawbacks that any inflationary scenario faces, thus motivating us to search for alternative solutions to the problems of Big Bang cosmology. In the reminder of
the paper we present such an alternative, based on a curved geometry.

\subsection{Mechanisms and Generic Predictions of Inflationary Models}\label{SS:PredictionsInfl}

Inflation is made possible whenever the universe is dominated by states of high energy density that do not dilute significantly with the cosmic expansion.  As
alluded before, one possibility for the high energy density state is due to curved space corrections to the energy-momentum tensor of a scalar field. This is the
case for what is currently known as $R^2$ or Starobinsky inflation~\cite{Starobinsky:1980,Starobinsky:1983,Kofman:1985}. In 1981 Alan Guth~\cite{Guth:1980} showed
how inflation can solve the horizon, flatness, and magnetic monopole problems of BB cosmology. In contrast to Starobinsky's model, in Guth's original proposal
inflation was driven by a scalar field (inflaton) trapped in a high energy, false vacuum, state. This mechanism, commonly referred to as ``old inflation'', was
soon shown to be unviable, as the phase transition to the true vacuum via bubble collisions cannot be completed efficiently and the universe continues to inflate
forever in most patches thus generating large inhomogeneities~\cite{Hawking:1982,Guth:1982}.

A solution to this ``graceful exit'' problem was found in 1982 by Linde~\cite{Linde:1981} and Albrecht \& Steinhardt~\cite{Albrecht:1982}, who proposed that
inflation is driven by a scalar field starting initially perched on the plateau of its effective potential. Subsequently the field ``slow rolls'' towards the
minimum (true vacuum state), with no quantum tunneling necessary. This ``new inflation'' scenario has its own theoretical drawbacks, especially when one attempts
to implement realizations of this mechanism based on the theory of high temperature phase transitions. One then has to postulate that the universe started
relatively homogeneous prior to inflation, therefore rendering one of the merits of the theory itself into a possible problem. Moreover, this scenario requires
existence of a pre-inflationary thermal state of the universe. This motivated introduction by Andrei Linde, in 1983, of ``chaotic
inflation''~\cite{Linde:Chaotic}; in this scenario inflation is driven by a scalar field being initially trapped, due to Hubble friction, at a high value of its
potential. In contrast to new inflation, chaotic inflation does not require assuming existence of an initial false vacuum state. This is replaced by the
assumption that the scalar field initial conditions are chaotic, and that at least in some patches of the universe the inflaton starts at high values
(``large-field''). Chaotic initial conditions can be useful even when inflation occurs near the maximum of the potential, such as in new inflation, ``low-field''
models. In fact they are more natural initial conditions, as opposed to the thermal equilibrium state required in the original version of new inflation. This
scenario is commonly referred to as ``hilltop inflation'' in the literature.

The next theoretical development was the realization that once started, inflation never completely ends everywhere in the universe, at least in most models. In
the case of plateau-like potentials of new inflation there is always a non-zero probability of finding the inflaton field in the decaying metastable false vacuum
state (see, e.g., Refs.~\refcite{StLin,Vilenkin:1983}). For chaotic inflation, large quantum fluctuations of the scalar field may end up kicking it back up the
potential, leading to a process of eternal self-reproduction of the universes, i.e. ``eternal inflation''~\cite{Linde:1986,Goncharov:1987}. On Hubble sized
patches the eternal inflationary universe appears homogeneous, however on much larger scales it has a fractal like structure, consisting of many bubble universes,
i.e. the ``multiverse''! This scenario raises yet unsolved problem of defining probabilities in an infinite multiverse, as the distinction between rare and
common events becomes ambiguous without some regularization scheme. We will discuss this in more detail in Sec.~\ref{SS:ProblemsInfl}. For recent reviews and
current status and perspectives on inflationary theory written by some of its original proponents see Refs.~\cite{Guth:2000,Ijjas:2013,Guth:2013,Linde:2014}

Measurements of the temperature of the CMB are uniform to one part in $10^5$, indicating that the Universe is homogeneous and isotropic to a high degree prior to
recombination. In standard BB cosmology this becomes problematic, since extrapolating back to the BB, assuming radiation domination, would lead to the universe at
last scattering being comprised of an extremely large number of causally disconnected patches. This is the ``horizon problem''. The accelerated expansion phase of
inflationary cosmology solves it, as now our observable universe can be easily generated from one single, causally connected patch that will be stretched out
during inflation. To get a sense of numbers, if the Universe starts to undergo inflation when it had a size corresponding to the Planck length, $l_P\sim10^{-33}$
cm, in $10^{-30}~\mathrm{s}$ of inflation it will become many orders of magnitude larger than our current observable universe ($l~\sim 10^{28}$ cm)! By the same
argument inflation solves all the other cosmological problems of BB. It leads to a Universe that is homogeneous on large scales and spatially flat, i.e.
$\Omega_K\equiv 1-\Omega\ll1$, since any initial amount inhomogeneity/curvature or any other unwanted relics is stretched out/diluted to unobservable levels
during the vacuum dominated expansion, if inflation lasts long enough.

Single field inflationary models typically predict Gaussian spectrum of scalar perturbations, in agreement with CMB data. Quantum sized fluctuations in the
inflaton field are stretched by the exponential expansion and become classical prior to sourcing the CMB temperature anisotropies. Therefore the Gaussian nature
of perturbations, as inferred from CMB temperature anisotropies, is due to the Gaussian statistics in the case of a single quantum field.  The power spectrum
contains all the statistical information needed in that case, as odd n-point correlation functions are identically zero and all the others can be related to
products of the two point correlation, i.e. to the power spectrum. In multi-field or single field models with non-trivial kinetic terms, or whenever the slow roll
conditions are violated one expects a significant amount of non-Gaussianity, even for the scalar modes.

In principle there are two different possible types of primordial perturbations: ``curvature'' (or adiabatic) and ``isocurvature''(or entropy). In the former case
each species has an equal perturbation in the number density, $\delta n_f/n_f$. For entropy perturbations one has $\delta\rho=0$, for the total fluid, therefore
the total density, or local curvature, remains homogeneous. Any generic primordial perturbation can be decomposed in a combination of those two orthogonal types.
Requiring that the primordial perturbations have an amplitude that allows them grow via gravitational instability to become the bound structures we observe today
on small scales (clusters, etc.) and matching this with the amplitude measured by CMB experiments at larger scales can place constraints on the type of primordial
perturbations, assuming scale invariance. If isocurvature perturbations were the sources for gravitational structures, then the anisotropy in the CMB would be
about 6 times larger than the one measured~\cite{Liddle:1993}. COBE was the first experiment to fix this amplitude and  since then adiabatic perturbations are
favored. As shown in Refs.~\refcite{Hawking:Irreg,Guth:Fluctuations,Bardeen:1983}, assuming radiation domination at the end of inflation, the comoving curvature
perturbation can be related to the inflationary potential $V$ and its derivative, $V_{\phi}=dV/d\phi$, in the following way:

\begin{equation}\label{Eq:CurvPert}
\mathcal{R}=-H\frac{\delta\phi}{\dot{\phi}}\sim\frac{H^2}{2\pi\dot{\phi}}=\frac{V^{3/2}}{2\sqrt{3}\pi V_{\phi}}
\end{equation}

During inflation both $H$ (the Hubble parameter) and $\dot{\phi}$ change very slowly, therefore one generic prediction is that the spectrum is nearly flat, i.e.
scale invariant\footnote{A careful analysis shows that there is a mild, logarithmic deviation from scale invariance. This general result was first discovered in
the context of Starobinsky inflation by Mukhanov \& Chibisov~\cite{Mukhanov:1981}}. Scalar perturbations generated during inflation become super-horizon and no
longer evolve whenever $k\sim aH$. At a later stage, after inflation ends, modes re-enter the horizon and evolve again, starting with the lower wavelength ones,
in a last-out first-in fashion. Gravitational waves, also known as tensor modes, can also be generated during inflation, as discussed in
Ref.~\refcite{Rubakov:GravWaves} for example. For massless graviton there are two independent polarizations $(h^+,h^\times)$ of the transverse and traceless parts
of the metric. It is convenient to introduce the power spectrum of perturbations in the following way:

\ba{Eq:PSpec}
\langle\mc{R}(\mb{k}_1)\mc{R}(\mb{k}_2)\rangle&=&(2\pi)^3\frac{2\pi^2}{k^3}\mc{P}_{\mc{R}}(k)\delta^3(\mb{k}_1+\mb{k}_2)\\
\langle h^{+,\times}(\mb{k}_1)h^{+,\times}(\mb{k}_2)\rangle&=&(2\pi)^3\frac{2\pi^2}{k^3}\mc{P}_{h^{+,\times}}(k)\delta^3(\mb{k}_1+\mb{k}_2), \ea with
$\langle...\rangle$ denoting ensemble average fluctuations. For the gravitational waves, we consider the sum of the two independent polarization and define the
tensor power spectrum as $\mc{P}_t=\mc{P}_{h^+}+\mc{P}_{h^\times}$. The scale dependence of the power spectrum is parametrized in the following way:

\ba{Eq:Ns}
\mc{P}_{\mc{R}}(k)&=&A_s(k_*)\left(\frac{k}{k_*}\right)^{n_s-1+\frac{1}{2}\alpha_s(k_*)\ln(k/k_*)+...}\\
\mc{P}_t(k)&=&A_t(k_*)\left(\frac{k}{k_*}\right)^{n_t+\frac{1}{2}\alpha_t(k_*)\ln(k/k_*)+...}, \ea where $A_{s,t}$ is the amplitude and $n_{s,t}$ is the spectral
index for the scalar and tensor modes respectively. Running of the spectral index is quantified by $\alpha_{s,t}\equiv dn_{s,t}/d\ln k$. In both cases $k_*$ is an
arbitrary reference or pivot scale where the normalization can in principle be fixed experimentally. In the slow roll approximation one can express the amplitudes
and spectral indices of the potential $V$ and its derivatives ($V_{\phi}=dV/d\phi$, $V_{\phi\phi}=d^2V/d\phi^2$) in the following way:

\begin{align}\label{Eq:SRoll}
    A_s &\approx \frac{V}{24\pi^2M_{pl}^4\epsilon}             & A_t &\approx \frac{2V}{3\pi^2M_{pl}^4}\\
    n_s-1 &\approx 2\eta-6\epsilon                             & n_t &\approx -2\epsilon,
\end{align}
with $\epsilon\equiv M_{pl}^2V_{\phi}^2/2V^2$ and $\eta=M_{pl}^2V_{\phi\phi}/V$ being parameters that in the slow roll regime are $\ll 1$. There are similar,
higher order expressions for the running of the spectral indices (see, e.g., Eqs.~(17)-(19) in Ref.~\refcite{Planck:xxii}). One can see from Eq.~\rf{Eq:SRoll}
that the scalar modes generated during a slow roll phase have a nearly scale invariant power spectrum, with deviations sensitive to features in the potential. The
value for the ratio of the tensor to scalar power spectra at the pivot scale can be obtained in the slow roll approximation by combining
Eqs.~\rf{Eq:Ns}-\rf{Eq:SRoll}: \be{Eq:ScalarTensor} r=\frac{\mc{P}_t(k_*)}{\mc{P}_{\mathcal{R}}(k_*)}\approx 16\epsilon\approx-8n_t \ee This important result is
known as the consistency relation, and in multi-field models becomes an inequality.

Last, but not least, inflaton perturbations naturally lead to a specific pattern in the spectrum of CMB radiation which has been confirmed experimentally by WMAP
and Planck satellites.  This provides an extremely powerful link that spans over $20$ orders of magnitude between the low energy Universe during recombination
($T\sim$ eV) and the high energy universe at the end of inflation ($T\sim 10^{15}$ GeV), made possible by the existence of large scale, super-horizon, modes that
did not evolve much after exiting the horizon during inflation. For a theoretical computation of the CMB spectrum of perturbations, that allows a transparent
understanding of the connections between the CMB features and basic cosmological parameters such as the spectral index and amplitude of the primordial
perturbations generated during inflation, see Ref.~\refcite{Mukhanov:CMBSlow}.

It is worth mentioning that sufficiently complicated models of inflation can deviate from any of the predictions mentioned in this section, even the ones taken
almost for granted such as the flatness of the universe!

\subsection{Tests and Current Status}\label{SS:TestsInfl}
CMB is the best available probe we have of the early Universe. Its temperature anisotropies and polarization measured with accuracy by the Planck satellite, the
Atacama Cosmology Telescope (ACT) and the South Pole Telescope (SPT) can be used to test predictions of inflation. First they fix the amplitude of the scalar
primordial fluctuations to $A_s=2.196^{+0.051}_{-0.06}\times10^{-9}$ and the scalar spectral index $n_s=0.9603\pm0.0073$ at the pivot scale $k_*=0.05$
Mpc$^{-1}$. This rules out exact scale invariance, and indicates a red tilt ($n_s<1$) of the spectrum. The amplitude measured from the CMB anisotropy has a value
that is compatible  to the one required in order to generate the observed structures in the late Universe such as galaxy clusters from adiabatic initial
perturbations! Moreover, Planck data does not indicate any non-Gaussianity, or isocurvature perturbation, favoring simple, single field models.

\begin{figure}[ph]
\centerline{\includegraphics[width = 10.0cm]{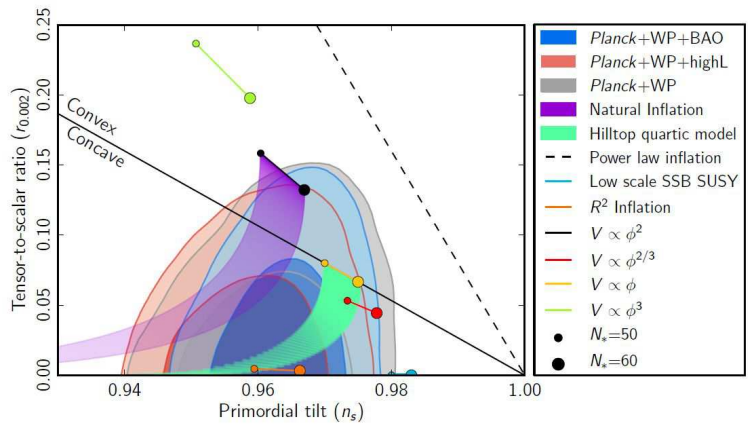}} \vspace*{8pt} \caption{Marginalized $68\%$ and $95\%$ CL regions for $n_s$ and $r$ from Planck and
other data sets: WP refers to the WMAP large-scale polarization likelihood, highL is high l data from ACT and SPT telescopes. Superimposed are predictions from
various inflationary models for $50-60$ e-foldings. [From Ref.~\cite{Planck:xxii}]\protect\label{Fig:nr}}
\end{figure}

The strongest constraints to date placed on inflationary models come from likelihood contours in the $n-r$ plane placed using a combination of datasets. As one
can see from Fig.~\ref{Fig:nr} concave, plateau like potentials are favored in addition to Hilltop, natural, Starobinsky and Higgs inflation models, which fit the
data well. More importantly, one can rule out a number of models, such as power-law (exponential potential), simplest hybrid models, chaotic models with monomial
potentials (i.e. $V\propto\phi^{p}$), if $p>2$. For an in depth analysis of many  of the surviving models see Refs.~\refcite{Encyclopedia,Best}.

In 2014 the BICEP2 experiment claimed detection of B-mode polarization in the CMB~\cite{BICEP2}, a pattern that can be generated by the tensor perturbations of
the metric produced during inflation. Their result implied that $r=0.2$, in strong tension with the bound from CMB data by Planck in combination with WMAP
polarization results: $r\lesssim 0.12$! Recently this issue has been settled, when a combined analysis of the data by the Planck and BICEP teams in
Ref.~\refcite{PlanckBICEP} lead to the conclusion that the signal can be attributed to large degree to galactic dust and placed an upper bound $r<0.12$. A
detection of tensor modes would have important consequences for inflationary theories. First that would fix the scale of the inflationary potential, as from
Eqs.~\rf{Eq:SRoll}-\rf{Eq:ScalarTensor} one gets:

\be{Eq:Potential} V^{1/4}=\left(\frac{r}{0.12}\right)^{1/4}1.9\times 10^{16}\mathrm{GeV} \ee Additionally a value of the tensor to scalar ratio $r\gtrsim 0.01$
would favor models with large, super Planckian, inflaton field excursions, as can be seen from the Lyth bound, a relation that relates the inflaton field
evolution to the number of e-folds, N, and $r$: \be{Lyth} \frac{\Delta \phi}{M_{pl}}\approx \frac{1}{\sqrt{8}}\int_0^N\,dN\sqrt{r} \ee $N$ is usually assumed to
be an integer between $50-60$ in most inflationary models.

In summary, basic predictions of single field inflation such as a spectrum of adiabatic, nearly scale invariant, Gaussian primordial perturbations have been
confirmed by the current CMB data. This has been interpreted by many as a very strong case for inflation. However, as we shall see in the next section, there are
severe theoretical unsolved problems that the inflationary paradigm faces.
\subsection{Theoretical Problems and alternatives to Inflation}\label{SS:ProblemsInfl}
In this section we briefly discuss some of the open questions of inflationary cosmology. For simplicity we will omit the trans-Planckian, $\eta$, and Higgs
instability problems, focusing on the more generic, model independent issues, that are most difficult to overcome.
\subsubsection{Reheating and subsequent expansion history}
At the end of inflation the inflaton field must decay into standard model particles, in a process called reheating. This has important observable consequences,
because in order to relate predictions of slow roll inflations to observables in the CMB one needs to know the time $t_*$ when the observable pivot scale has
exited the horizon during inflation. Obviously this depends on the details of reheating and the subsequent expansion of the Universe. Neither of those two are yet
fully constrained. Reheating can proceed in a number of different ways, and one has to take this into account when deriving constraints in parameter space, as
done for instance in Ref.~\refcite{Reheating}. The expansion history of the universe is constrained by the requirement that radiation domination starts prior to
the Big Bang Nucleosynthesis era, or when the temperature of the universe was of the order of $\sim$MeV. It is natural to assume that radiation domination starts
as soon as inflation ends, as the inflaton will decay into light species. However, during reheating in a quadratic potential, the coherent oscillations of the
inflaton field make it act as pressureless dust. Depending on how slow the inflaton field decays into radiation, this ``early matter dominated era'' (EMDA) can
have important consequences on the expansion history, and thus on relating observables measured at CMB scales to inflationary theory.

\subsubsection{Initial Conditions}
In order for inflation to start one needs a Universe homogeneous on scales larger than the Hubble radius, prior to inflation, as large kinetic and gradient terms
inhibit inflation. Thus, in order to solve the BB cosmology homogeneity and flatness problem one needs to start from a very homogeneous patch. One ``natural''
solution to this problem, proposed by Linde in Ref.~\refcite{Linde:Chaotic}, is to assume chaotic initial conditions: by the time the Universe reached the Planck
energy scale, all different energy densities are of the same order. In the case of single field inflation this amounts to:
$\frac{1}{2}\dot{\phi}^2\sim\half(\partial_i\phi)^2\sim V(\phi)\sim M_{pl}^4$. The potential term will quickly dominate the energy density and therefore inflation
can proceed. Patches which start with large kinetic or gradient terms compared with the potential do not inflate, and therefore are disfavored at a classical
level. However, CMB data, and the upper bound it places on the tensor modes: $r\lesssim 0.12$, implies that the scale of inflaton potential is much smaller than
the Planck energy: $V\lesssim 10^{-12}M_{pl}^4$, as one can see from Eq.~\rf{Eq:Potential}.  As argued in Ref.~\refcite{Ijjas:2013}, for the case of featureless,
plateau-like potentials, such as the ones favored by the data, this becomes problematic. First, this class of potentials require an amount of fine tuning much
larger than the now-disfavored power-law potentials. More importantly, extrapolating back to the Planck scale one finds initial conditions where gradient,
inflation prohibitive, terms dominate over the potential. In order for inflation to proceed one needs to start from a patch that is homogeneous on scales $1000$
times larger than the Hubble horizon! A possible solution to this problem, postulating that inflation starts from a region of negative spatial curvature, has been
proposed in Ref.~\refcite{Guth:2013}. This assumption can reduce the required homogeneity length at Planck scale from $\sim 1000$ to $\sim 1-15$ Hubble lengths.
Features in the potential at scales not probed via CMB anisotropies, and hence of no observational consequence, could also help to solve this initial homogeneity
problem. One possibility is to assume an initial stage of inflation that proceeds just as in the case of old inflation, via tunneling from a false vacuum and
bubble nucleation. Symmetry of the true vacuum bubbles guarantee homogeneity prior to the onset of the last stage of inflation. A very intricate possibility that
seems outside of experimental reach, and that is introduced only to reduce the homogeneity scale.

\subsubsection{Multiverse and the measure problem}
For inflationary models with plateau-like potentials, such as the ones favored by CMB data, there are regions where the quantum fluctuations over a Hubble time,
$\Delta\phi_{qu}\sim H/2\pi$ dominate over the classical evolution, $\Delta\phi_{cl}=\dot{\phi}/H$. As explained in Sec.~\ref{SS:PredictionsInfl}, this leads to
an eternally self-reproducing regime, where inflation never ends globally and the universe has a fractal-like structure on scales much larger than the Hubble
length. Actually, eternal inflation and the multiverse is a very generic picture, and there are very few models that do not have a self-reproducing regime. This
leads to a loss of predictability, as any cosmological possibilities are realized in at least some of the bubble universes that are part of the multiverse.
Probabilities cannot be defined unless a regularization scheme is used, since in an infinite multiverse they involve comparison on infinities\footnote{For a
review of the measure problem in the multiverse see e.g. Ref.~\cite{Guth:2000}}. For example Refs.~\refcite{Bousso:2006,Bousso:2008,Nomura:2011} propose such
regularization schemes; however the predicted probabilities are scheme dependent, and there is no consensus yet as to which one is correct. A na\"{i}ve, weight by
volume, measure will actually exponentially favor a much younger patch than our Universe~\cite{Guth:2000} or Boltzmann brains~\cite{Albrecht:2004}. Those are
known as the youngness paradox and Boltzmann brain problem, respectively.

\subsubsection{Large scale correlations and structure problems}
\begin{figure}
\centerline{\includegraphics[width=8cm]{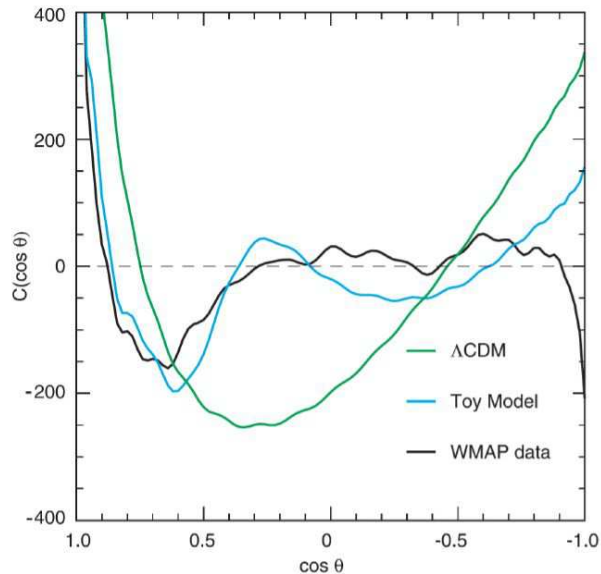}} \vspace*{8pt} \caption{\label{CMBCorrelations} Angular correlation function of the best fit
$\Lambda$CDM model, a finite size universe model, and WMAP data on large angular scales (adopted from Ref.~\cite{CMBWMAP1}).}
\end{figure}

Even if CMB data seems to confirm the vanilla $\Lambda$CDM paradigm, there are certain anomalies at large scales, indicative of an incomplete understanding of the
physics in the early universe. For instance, inflation is not consistent with observed large scale angular correlations in CMB data. Inflation models require
angular correlation at all angles, not only at angles up to $\sim 60^{\circ}$, because inflation occurred at all scales.  The discrepancy in angular correlations
between CMB and $\Lambda$CDM model that is presented in Fig.~\ref{CMBCorrelations} was first noticed in Ref.~\refcite{CMBCOBE} and confirmed later in
Refs.~\refcite{CMBWMAP1,SevenWMAP}, and ~\refcite{NineWMAP}.

\begin{figure}
\centerline{\includegraphics[width=8cm]{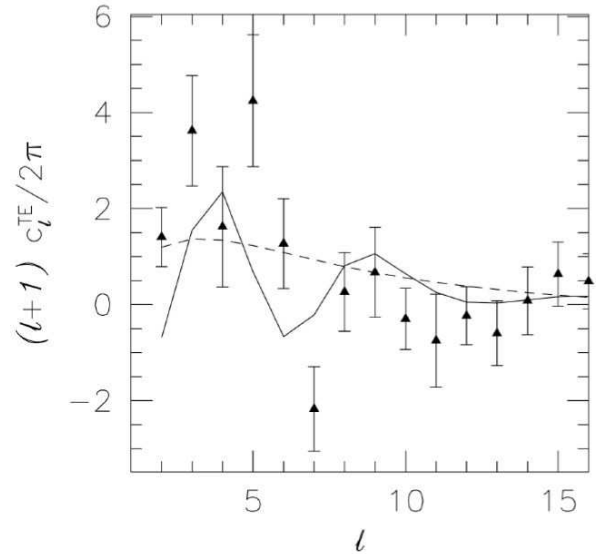}} \vspace*{8pt} \caption{\label{CMBTE} The comparison of the data to the predicted TE power spectrum in a
finite universe model (solid line) and the $\Lambda$CDM model (dashed line), adopted from Ref.~\cite{CMBWMAP1}.}
\end{figure}

There is an obvious difference between the CMB spectrum and predictions of the standard model. The figure also includes a curve that shows very good agreement
between the observable data and a finite size universe model (similar to the model proposed here). Please note that the finite size model gives not only a better
match to the observed correlation function than the  $\Lambda$CDM model, but also predicts the distinctive signature in the temperature polarization $(TE)$
spectrum; see Fig.~\ref{CMBTE}.

\subsubsection{Alternatives to inflation}
Motivated by the thorny theoretical problems of inflationary cosmology some authors have proposed alternative scenarios such as string gas cosmology, matter
bounces, ekpyrotic/cyclic scenarios (for a review, see e.g. Ref.~\refcite{Alternatives}). As shown in Ref.~\refcite{Gratton:2003} there are two robust cases for
the effective equation of state parameter $w$ required for a scalar field to produce a nearly scale-invariant spectrum of density perturbations: $w\approx -1$
(inflation) and $w\gg 1$ (ekpyrotic/cyclic). The latter scenario predicts almost no tensor perturbations and a degree of non-Gaussianity larger than simple,
single field inflation. If B-modes are detected in the CMB with future experiments, one could use the consistency relation in Eq.~\rf{Eq:ScalarTensor} to test the
slow roll hypothesis vs alternatives. A very interesting review of the ekpyrotic/cyclic scenario and its problems can be found in the Appendix of
Ref.~\refcite{Linde:2014}.

Of the three problems of BB cosmology the horizon problem is the most serious one, as ``there are possible solutions of the flatness and monopole problems that do
not rely on inflation''\cite{Weinberg:Cosmology}. The explanation for flatness may be the anthropic principle\cite{Barrow}, that intelligent life would only arise
in those patches of universe with $\Omega$ very close to 1; another explanation could be that space is precisely flat at time of BB. Guth's monopoles may be
explained by inflation, or the physics may be such that they never existed in appreciable abundances. An explanation may be that there is no simple gauge group
that is spontaneously broken to the gauge group $SU(3)\times SU(2)\times U(1)$ of the Standard Model. In the remainder of this paper we will discuss a novel,
geometric, approach that could be used to solve the horizon problem without the need for postulating an inflationary epoch.

\section{Positive curvature and $\omega=-1/3$ quintessence}

\setcounter{equation}{0}

Inflation explains uniformity of CMB and solves horizon problem by generating our observable universe from one single causally connected patch.
Homogeneity in the CMB on the level of $10^{-5}$ is explained by inflation era. However, arguments in favor of inflation only exist if space was already
homogeneous before inflation. If the pre inflationary universe was not already homogenous, inflation will not lead to homogeneity \cite{Goldwirth}.
So, the homogeneity problem is pushed only back in time, because the Big Bang itself is taken to be inherently free of correlations.

In addition, after the time when inflation ended to the moment of the last recombination, when CMB was emitted,  densities changed from $10^{38} ~\mathrm{kg/m}^3$
to $10^{-17} ~\mathrm{kg/m}^3$, and temperature from $10^{29}$ K to $3000$ K. The high degree of isotropy observed in the microwave background indicates that any
density variations from one region of space to another at the time of decoupling must have been small, at most a few parts in $10^{-5}$. This requires that
changes in density at any part of the universe is the same to the 60 orders of magnitude. After inflation ended, during the time period of 380k years some parts
of the universe are not anymore causally connected and there is no reason that they will have the same density and the same temperature at the time of decoupling.
Taking in account that during that period we also have acoustic oscillations (heat of photon-matter interactions creates a large amount of outward pressure that
counteract gravity) it is statistically unlikely that the CMB will have observed uniformity. This is similar to the horizon problem, but after inflation,
inflation does not help to solve it.

It is important to note that the transfer of the quantum fluctuations in the inflaton field to density perturbations that lead to the CMB anisotropies is not well
understood.
For instance, at the end of inflation during reheating the energy of vacuum was transferred to ordinary matter and radiation, but we have no clear idea how the
energy transfer took place and which particles are first created. Later, we can only guess when and how some particles effectively stopped interacting with the
rest of matter and radiation and become cold dark matter.  These uncertainties do not allow us to make predictions that will be accurate on 60 decimal places. So,
how do we explain the CMB uniformity?

We argue that the observed uniformity in the CMB does not mean that space was uniform at the time of decoupling. We propose a cosmological model that allows for a
different interpretation of the CMB data and for inhomogeneity of the universe at the early stage. A large-scale homogeneity and isotropy is not required by
classical GR theory. It is well known that in the Big Bang models homogeneity of space cannot be explained, it is simply assumed in initial conditions.

In our positive curvature closed universe model the CMB is always coming from a very small vicinity of the antipodal point. Therefore measuring the same CMB by
looking in the opposite directions of the universe does not represent or reflect the uniformity of the universe at the time of decoupling, because we always
measure CMB originating from approximately the same antipodal point regardless of the direction of observation. For that reason we always must obtain the same
result. Small variations for the CMB are possible and observed, but these variations are the result of measuring CMB from a small region and not exactly from a
single point, and because of the interaction between matter and light during its travel. For instance, depending on the direction we choose to measure the CMB,
light will travel through different galaxies and will interact with different amounts of matter, which will result in small observed variations of the CMB at
large angular scales (as the photons pass through large scale structures) by the integrated Sachs -- Wolfe effect \cite{4a,4b,4c,4d}.

To establish a connection between the uniformity of the earlier universe at the time of decoupling and the CMB we will need to make a completely different kind of
measurement of the CMB. We can see the CMB in any direction we can look in the sky. However, we must keep in mind that the CMB emitted from over-dense regions that would
ultimately form, for instance the Milky Way, is long gone. It left our part of the universe at the speed of light billions of years ago and now forms the CMB for
observers in remote parts of the universe, for an observer located at the antipodal point. To measure the uniformity of the universe at the time of decoupling we
will need to measure the CMB in at least two different points. If, that measurements give the same result, then and only then may we speak about the uniformity of
the CMB and uniformity of the universe at the time of decoupling. However, such measurements are not possible at the present time since we cannot move to
different place to perform such measurement.

Following the above and ideas expressed in the recent papers~\cite{BranislavTurin,Branislav}, let us try to find a possibility for an affirmative answer to the
following question: ``Is there a chance from the mathematical point of view that the last scattering surface is approximately point-like, or, in other words, that
the CMB radiation originates from a very limited space region in the vicinity of the only one point?" For this purpose, taking into account that the answer is
definitely negative in the framework of the conventional $\Lambda$CDM cosmological model with flat spatial geometry, we consider its extension with the positive
curvature, i.e. the closed space. The corresponding FLRW metric reads:
\be{1} ds^2=c^2dt^2-a^2(t)\left[d\chi^2+\sin^2\chi\left(d\theta^2+\sin^2\theta d\varphi^2\right)\right]\, ,\ee
where the hyperspherical coordinates $\chi\in[0,\pi]$, $\theta\in[0,\pi]$, $\varphi\in[0,2\pi)$; $c$ represents the speed of light, while $a(t)$ stands for the
scale factor. This function satisfies the well-known first Friedmann equation:
\ba{2} H^2&=&\left(\frac{\dot a}{a}\right)^2=\frac{\kappa\varepsilon_{rad}c^2}{3}+\frac{\kappa\bar\rho c^4}{3a^3}+\frac{\Lambda c^2}{3}-\frac{c^2}{a^2}\nn\\
&=&H_0^2\left[\Omega_{rad}\left(\frac{a_0}{a}\right)^4+\Omega_{mat}\left(\frac{a_0}{a}\right)^3+\Omega_{\Lambda}+\Omega_{K}\left(\frac{a_0}{a}\right)^2\right]\,
,\ea
where the dot denotes the derivative with respect to time $t$; $\kappa=8\pi G_N/c^4$, with $G_N$ being the Newtonian gravitational constant; $a_0$ is the current
scale factor value, and $H_0\approx67.4$ km/s/Mpc is the current value of the Hubble parameter $H(t)=\dot a/a$. Further, $\varepsilon_{rad}$ and $\bar\rho
c^2/a^3$ represent the energy densities of radiation and nonrelativistic matter (with the average rest mass density $\bar\rho$ in the comoving coordinates),
respectively, while $\Lambda$ stands for the cosmological constant. Finally, the following well-known energy fractions are introduced:
\ba{3} &{}&\Omega_{rad}=\frac{\kappa\varepsilon_{rad(0)}c^2}{3H_0^2}\approx0.000055,\quad \Omega_{mat}=\frac{\kappa\bar\rho c^4}{3H_0^2a_0^3}\approx0.31\, ,\nn\\
&{}&\Omega_{\Lambda}=\frac{\Lambda c^2}{3H_0^2}\approx0.69,\quad \Omega_{K}=-\frac{c^2}{H_0^2a_0^2}\sim-0.005\, ,\ea
where $\varepsilon_{rad(0)}$ denotes the today's radiation energy density. Here the numerical values approximately correspond to what is observed according to the
recent Planck results (see Ref.~\refcite{Planck}). It should be noted that the value $0.005$ may be considered as an approximate upper limit for the quantity
$|\Omega_K|$ in the considered case of the positive curvature space.

Introducing the dimensionless quantities $\tilde a=a/a_0$ and $\tilde t=H_0t$ (so $\tilde a(0)=1$ and $\tilde a(-\tilde t_0)=0$, where $\tilde t_0$ represents the
dimensionless age of the Universe), we get
\be{4} \tilde t_0=\int\limits_0^1\frac{\tilde ad\tilde a}{\sqrt{\Omega_{rad}+\Omega_{mat}\tilde a+\Omega_{\Lambda}\tilde a^4+\Omega_K\tilde a^2}}\approx0.96\,
,\ee
and this value corresponds to $t_0\approx13.9$ billions of years of the Universe evolution. In addition, it is worth mentioning that the today's value of the
deceleration parameter $q=-\ddot a/\left(aH^2\right)$ reads $q_0\approx-0.535$, being in complete agreement with the supernovae data.

Now, let us demand that
\be{5} \int\limits_{-t_0}^0\frac{cdt}{a(t)}\approx\pi\, .\ee

The physical interpretation of this condition is clear: if it holds true, than light travels between the antipodal points during the age of the Universe. One can
easily verify that it is impossible to reach the approximate equality \rf{5} within the standard pure $\Lambda$CDM model. Really, with the help of the values
\rf{3} for the left-hand side we immediately obtain
\be{6} \sqrt{-\Omega_K}\int\limits_0^1\frac{d\tilde a}{\sqrt{\Omega_{rad}+\Omega_{mat}\tilde a+\Omega_{\Lambda}\tilde a^4+\Omega_K\tilde a^2}}\approx0.2\neq\pi\,
.\ee

However, without the crucial condition \rf{5} the proposed elegant geometrical solution of the horizon problem is not valid. Therefore, in what follows we
continue demanding its fulfilment. Of course, this is apparently forbidden if the composition of the Universe remains
unchanged. In this connection we supplement the positive spatial curvature extension of the conventional cosmological model with an additional perfect fluid with
the constant parameter $\omega$ in the linear equation of state $p_Q=\omega \varepsilon_Q$, where $\varepsilon_Q$ and $p_Q$ represent its energy density and
pressure, respectively. If $-1<\omega<0$, such a perfect fluid may be called quintessence~\cite{quintess1,quintess2,quintess3}.

According to Ref.~\refcite{BUZ1}, only two negative values of the constant parameter $\omega$ are admissible from the point of view of the cosmological
perturbations theory: $\omega=-1$ (this possibility is already completely exhausted by introducing the nonzero cosmological constant $\Lambda$, which, as it is
known, can be interpreted as a perfect fluid with the vacuum equation of state $p_{\Lambda}=-\varepsilon_{\Lambda}$) and $\omega=-1/3$ (for the foundations of the
mechanical/discrete approach to cosmological problems inside the cell of uniformity, leading directly to these severe theoretical restrictions, see
Refs.~\refcite{EZcosm1,EKZ2,EZcosm2}). Consequently, we make the choice $\omega=-1/3$. It is interesting that the frustrated network of such topological defects
as cosmic strings~\cite{ShellVil,Kumar} is characterized by exactly the same value of the parameter $\omega$. Such a constituent is also used in
Refs.~\refcite{Melia1,Melia2} within another alternative cosmological model.

It should be noted that the extension of the standard $\Lambda$CDM model with respect to the positive curvature space without quintessence has an important
problematic aspect: the gravitational potential $\phi$ in the comoving coordinates produced by a point-like mass $m$ diverges at the antipodal point where this
mass is actually absent, and this may be considered as a disadvantage of the investigated spherical topology.  As shown in Ref.~\refcite{EZcosm1}, if
the considered gravitating mass is situated at the point $\chi=0$, then
\be{7} \phi=2C\cos\chi-G_Nm\left(\frac{1}{\sin\chi}-2\sin\chi\right)\, ,\ee
where $C$ is some integration constant. It immediately follows from \rf{7} that the function $\phi$ diverges not only at the point $\chi=0$ where it has the
correct Newtonian limit $\phi\rightarrow-G_Nm/\chi$, as it certainly should be, but also at the antipodal point $\chi=\pi$ where there is no any mass! Let us note
as well that the inevitably coming to mind conclusion that this result is nonphysical may be drawn in the case of the infinite-range gravitational interaction
analyzed in Ref.~\refcite{EZcosm1}. At the same time, in the opposite case of the finite-range gravitational interaction studied in detail in Ref.~\refcite{EBV}
the situation may improve drastically. However, this chance lies beyond the scope of the present manuscript.

The situation with the gravitational potential improves as well, if one introduces the above-mentioned $\omega=-1/3$ quintessence Ref.~\refcite{BUZ1}. Now under
the same problem statement
\be{8} \phi=-G_Nm\frac{\sin\left[(\pi-\chi)\sqrt{\mu^2+1}\right]}{\sin\left(\pi\sqrt{\mu^2+1}\right)\sin\chi}\, ,\ee
where $\mu^2=(3-\kappa\varepsilon_{Q(0)}a_0^2)>0$. Here, in its turn, $\varepsilon_{Q(0)}$ denotes the today's quintessence energy density. Obviously, for
$\sqrt{\mu^2+1}\neq2,3,\ldots$ the function $\phi$ \rf{8} is finite at any point $\chi\in(0,\pi]$, including the suspect antipodal point $\chi=\pi$. The
finiteness remains in force also in the opposite case $(3-\kappa\varepsilon_{Q(0)}a_0^2)<0$.

Avoidance of the gravitational potential divergency in the presence of quintessence in the closed Universe may serve as a valid reason to incorporate such a
constituent into the cosmological model under consideration. The other reason consists in allowing the condition \rf{5} of light traveling between the antipodal
points during the age of the Universe to be satisfied. The next section is entirely devoted to determination of numerical values of the corresponding energy
fractions.

\section{Comparison with supernovae data}

\setcounter{equation}{0}

The contribution of quintessence to the right-hand side of the first Friedmann equation \rf{2} reads:
\be{9} \frac{\kappa\varepsilon_{Q}c^2}{3}=H_0^2\Omega_Q\left(\frac{a_0}{a}\right)^2,\quad \Omega_Q=\frac{\kappa\varepsilon_{Q(0)}c^2}{3H_0^2}\, .\ee

Neglecting the radiation contribution, we get
\be{10} \Omega_{mat}+\Omega_{\Lambda}+\Omega_K+\Omega_Q=1\, .\ee

Demanding that the acceleration parameter $q$ has approximately the same current value as in the standard $\Lambda$CDM model, that is $q_0\approx-0.535$, we
obtain one more equation for the energy fractions:
\be{11} -\frac{1}{2}\Omega_{mat}+\Omega_{\Lambda}\approx0.535\, .\ee

Besides, instead of \rf{6} we have now
\be{12} \sqrt{-\Omega_K}\int\limits_0^1\frac{d\tilde a}{\sqrt{\Omega_{mat}\tilde a+\Omega_{\Lambda}\tilde a^4+\Omega_K\tilde a^2+\Omega_Q\tilde a^2}}\approx\pi\,
.\ee

For illustrative purposes we single out two concrete examples. First, it is the so-called ``exact compensation" case when the positive contribution $\Omega_Q$ of
quintessence exactly compensates the negative one of the spatial curvature $\Omega_K=-|\Omega_K|$: $\Omega_Q=-\Omega_K\approx0.93$, so the rest two fractions
exactly coincide with those from the conventional model, namely $\Omega_{mat}\approx0.31$, $\Omega_{\Lambda}\approx0.69$. In this case the gravitational potential
in the comoving coordinates produced by a point-like mass reads:
\be{13} \phi=\frac{G_Nm}{2\pi}-G_Nm\frac{\cos\chi}{\sin\chi}\left(1-\frac{\chi}{\pi}\right)\, ,\ee
again being a finite function of $\chi\in(0,\pi]$.

Second, it is the so-called ``visible matter" case when the nonrelativistic matter contribution equals $\Omega_{mat}\approx0.040$ (which allows interpreting these
$4\%$ as approximately corresponding to the visible matter only, without dark matter). Then $\Omega_{\Lambda}\approx0.555$, $\Omega_Q\approx 0.721$ and
$\Omega_K\approx-0.316$. Both found sets of energy fractions satisfy the equations \rf{10} and \rf{11} as well as the condition \rf{12}. In the following two
tables we present some useful calculations for the ``exact compensation" and ``visible matter" cases, respectively.

\begin{table}[h]
\tbl{Numerical values for the ``exact compensation" case.}
{\begin{tabular}{@{}cccccc@{}} \toprule
{$\ z\ $} & {$\ a/a_0\ $} &  {$\ -t\ $} & {$\ -ct\ $} & {$\ d_L\ $} &
{$\ d_{L(\Lambda CDM)}\ $} \\
&  & (Gyr) & (Gpc) & (Gpc) & (Gpc)\\
\colrule
0.1\hphantom{00} & \hphantom{0}0.91 & \hphantom{0}1.34 & \hphantom{0}0.41 & \hphantom{0}0.48 & \hphantom{0}0.48  \\
0.5\hphantom{00} & \hphantom{0}0.67 & \hphantom{0}5.17 & \hphantom{0}1.59 & \hphantom{0}2.85 & \hphantom{0}2.94  \\
1.0\hphantom{00} & \hphantom{0}0.50 & \hphantom{0}7.99 & \hphantom{0}2.45 & \hphantom{0}6.23 & \hphantom{0}6.84  \\
1.5\hphantom{00} & \hphantom{0}0.40 & \hphantom{0}9.58 & \hphantom{0}2.94 & \hphantom{0}9.57 & \hphantom{0}11.27 \\
2.0\hphantom{00} & \hphantom{0}0.33 & \hphantom{0}10.63 & \hphantom{0}3.26 & \hphantom{0}12.69 & \hphantom{0}16.04  \\
1000\hphantom{00} & \hphantom{0}0.001 & \hphantom{0}13.89 & \hphantom{0}4.26 & \hphantom{0}489.4 & \hphantom{0}14046  \\ \botrule
\end{tabular}\label{Tab:exactcomp} }
\end{table}

\begin{table}[h]
\tbl{ Numerical values for the ``visible matter" case.}
{\begin{tabular}{@{}cccccc@{}} \toprule
{$\ z\ $} & {$\ a/a_0\ $} &  {$\ -t\ $} & {$\ -ct\ $} & {$\ d_L\ $} &
{$\ d_{L(\Lambda CDM)}\ $} \\
&  & (Gyr) & (Gpc) & (Gpc) & (Gpc)\\
\colrule
0.1\hphantom{00} & \hphantom{0}0.91 & \hphantom{0}1.34 & \hphantom{0}0.41 & \hphantom{0}0.48 & \hphantom{0}0.48  \\
0.5\hphantom{00} & \hphantom{0}0.67 & \hphantom{0}5.23 & \hphantom{0}1.61 & \hphantom{0}2.95 & \hphantom{0}2.94  \\
1.0\hphantom{00} & \hphantom{0}0.50 & \hphantom{0}8.26 & \hphantom{0}2.54 & \hphantom{0}6.89 & \hphantom{0}6.84  \\
1.5\hphantom{00} & \hphantom{0}0.40 & \hphantom{0}10.13 & \hphantom{0}3.11 & \hphantom{0}11.36 & \hphantom{0}11.27 \\
2.0\hphantom{00} & \hphantom{0}0.33 & \hphantom{0}11.45 & \hphantom{0}3.51 & \hphantom{0}16.11 & \hphantom{0}16.04  \\
1000\hphantom{00} & \hphantom{0}0.001 & \hphantom{0}16.89 & \hphantom{0}5.17 & \hphantom{0}1385.7 & \hphantom{0}14046  \\ \botrule
\end{tabular}\label{Tab:vismatter} }
\end{table}

\begin{figure}[htb]
\centerline{\includegraphics[width=8cm]{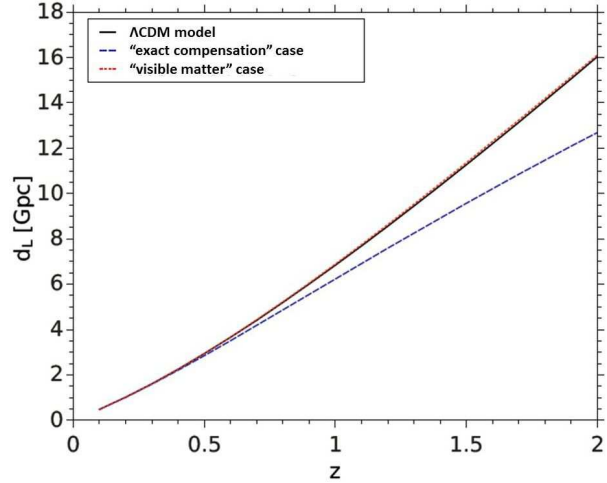}}
\vspace*{8pt}
\caption{The dependence of the luminosity distance $d_L$ on the cosmological redshift $z$ for the standard $\Lambda$CDM model (a black solid line)
and both the ``exact compensation" (a blue dashed line) and ``visible matter" (a red dotted line) cases.\label{Fig:LD}}
\end{figure}

The last two columns of these tables show the numerical values of the luminosity distance
\be{14} d_L(z)=\frac{c}{H_0}(1+z)\sqrt{-\frac{1}{\Omega_K}}\sin\left[\sqrt{-\Omega_K}\int\limits_{\frac{1}{1+z}}^1\frac{d\tilde a}{\sqrt{\Omega_{mat}\tilde
a+\Omega_{\Lambda}\tilde a^4+(\Omega_Q+\Omega_K)\tilde a^2}}\right]\ee
within the cosmological model under consideration and the luminosity distance
\be{15} d_{L(\Lambda CDM)}(z)=\frac{c}{H_0}(1+z)\int\limits_{\frac{1}{1+z}}^1\frac{d\tilde a}{\sqrt{\Omega_{mat}\tilde a+\Omega_{\Lambda}\tilde a^4}}\ee
in the framework of the standard $\Lambda$CDM one, respectively, as functions of the cosmological redshift $z=a_0/a-1$. These dependences are also depicted in
Fig.~\ref{Fig:LD}. We see that both examples agree with the supernovae data quite well for sufficiently small $z$. Nevertheless, for large $z$ the discrepancy
between the considered model and the conventional one with respect to the luminosity distance becomes wide. Indeed, for $z_*=1100$ (this redshift corresponds
approximately to the recombination time) we have $d_{L(\Lambda CDM)}\approx15400$ Gpc while for two above-mentioned sets of energy fractions the luminosity
distance is about $510$ Gpc or $1450$ Gpc, respectively. This leads to drawbacks discussed briefly in the next section.

\section{Cosmic microwave background anisotropy}

\setcounter{equation}{0}

Certainly, it is not enough to reach agreement only with the supernovae data. Any novel cosmological scenario must take a lot of other tests in order to be called
successful and viable. Among them here we single out consistency with the CMB data. The simplest possible check consists in estimation of the location of the
first acoustic peak directly related to the so-called acoustic scale (see, e.g., Refs.~\refcite{PlanckCMB,Ruth,Rubakov}):
\be{theta} \theta_{*}=\frac{r_s}{D_A}\, ,\ee
representing the ratio of the comoving size of the sound horizon at the time of recombination and the corresponding angular diameter distance, which is directly
proportional to the luminosity distance $d_L(z_*)$ divided by $(1+z_*)^2$. Let us restrict ourselves, for example, to the ``visible matter" case, which matches
the luminosity distance redshift relation of $\Lambda$CDM to within $1\%$ up to high redshifts, as one can see from Fig.~\ref{Fig:LD}. Then from the quite natural
requirement $\theta_*/\theta_{*(\Lambda CDM)}\sim 1$ we obtain:
\be{16} \frac{r_s}{r_{s(\Lambda CDM)}}\sim \frac{D_A}{D_{A(\Lambda CDM)}}=\frac{d_L}{d_{L(\Lambda CDM)}}\approx\frac{1450}{15400}\approx0.1\, .\ee

Unfortunately, it can hardly be so. Indeed, since the comoving size of the sound horizon $r_s$ is determined by the average sound speed before the recombination,
the requirement \rf{16} means that in the framework of the alternative cosmological model under consideration the average sound speed during this earliest epoch
of the Universe evolution should be decreased in about $10$ times as compared to the standard $\Lambda$CDM one. However, this quantity in its turn is determined
mainly by the radiation contribution and thus is approximately equal to $1/\sqrt{3}$ if defined as $\sqrt{\partial p/\partial\varepsilon}$. The physical mechanism
of such decreasing of the sound speed is actually completely unknown to us. The reasoning may be as follows. First, we cannot simply compensate the radiation
contribution to the sound speed squared by some other contribution of the opposite (negative) sign, because all popular exotic perfect fluids including
quintessence become relativistic at the earliest evolution stage if their initial peculiar velocities are nonzero, and therefore are described by the same
parameter $1/3$ in the equation of state as the radiation. Second, even if there is no thermal motion of some assumed additional fluid at the very beginning of
the Universe evolution, which is hard to imagine, then this fluid should affect not only the sound propagation making it difficult, but also possess the energy
density and pressure compared to those of radiation. This would evidently have grave consequences not only for the epoch before the recombination but also for the
epoch after it, and from the experimental point of view this is also hard to imagine with respect to what is observed or supposed to be known about the early
evolution stages.

The other possibility to save the situation lies in the following. One can reject comparing the acoustic scales and change the initial power spectrum instead in
such a way that its new shape already contains the observed acoustic peaks. Then their positions may be adjusted to the experimental data. According,
e.g., to Ref.~\refcite{Ruth}, the conversion from the flat geometry to the spherical space (with the positive spatial curvature) may be implemented, in
particular, by replacing the today's conformal time $t_0$ by the comoving angular diameter distance $\chi(t_0)$ to the last scattering surface, where we have
switched for convenience to the corresponding designations adopted in Ref.~\refcite{Ruth}, see, e.g., the formulas (2.254) and (2.255), respectively, for the
positions of the peaks:
\be{} l_n\sim n\pi\sqrt{3}\frac{t_0}{t_*}\ (\Omega_{K}=0)\quad \rightarrow \quad l_n\sim n\pi\sqrt{3}\frac{\chi(t_0)}{t_*}\ (\Omega_K<0)\, .\ee

The difference between these quantities (see our estimation \rf{16}) leads to a conclusion that the repetition frequency for the peaks within the alternative
cosmological model under consideration would be theoretically much higher than in the conventional one, in disagreement with the experimental data, if the same
initial power spectrum is assumed, namely
\be{17} k^3P(k)\sim(kt_0)^{n-1}\, ,\ee
where, as usual, $n$ denotes the spectral index Ref.~\refcite{Ruth}. Note that, up to numerical factors, $k^3P(k)$ is what we have labeled
$\mathcal{P}_{\mathcal{R}}$ in Eq.~\rf{Eq:PSpec}. Moreover, we neglect running of the spectral index, as Planck data suggest no evidence for it, and assume here a
pivot scale equal to the horizon scale today, i.e. $k_*=1/t_0$. Now, in order to adjust the acoustic peaks of the CMB anisotropy, we are forced to assume the
other form of the spectrum, namely
\be{18} \tilde P(k)=\left(\frac{\chi(t_0)}{t_0}\right)^3P\left(k\frac{\chi(t_0)}{t_0}\right)\, ,\ee
so that
\be{19} k^3\tilde
P(k)=\left(k\frac{\chi(t_0)}{t_0}\right)^3P\left(k\frac{\chi(t_0)}{t_0}\right)\sim\left(k\frac{\chi(t_0)}{t_0}t_0\right)^{n-1}=[k\chi(t_0)]^{n-1}\, .\ee

In other words, we absorb the difference between the angular diameter distance and the comoving distance for a curved universe in the amplitude of the
primordial power spectrum.  Let us estimate the change needed in order to match the location of the acoustic peaks in the CMB. From Eqs.~\rf{16},~\rf{17}
and~\rf{19} we get:

\be{Eq:Ratio}
\frac{\tilde P(k)}{P(k)}=\left(\frac{\chi(t_0)}{t_0}\right)^{n-1}=\frac{\tilde A_s(k_*)}{A_s(k_*)}=(0.1)^{n-1}\stackrel{n=0.96}{\Longrightarrow}1.096
\ee

Therefore, if one assumes $n=0.96$, the best fit value of the spectral index of scalar perturbations from CMB, then the change in the amplitude of the power
spectrum needs to be of the order of $10\%$, to compensate for a $90\%$ difference in the location of the first acoustic peak. This drastic reduction is made
possible by the near scale invariance! CMB fixes the amplitude of the scalar modes to $A_s=2.196^{+0.051}_{-0.06}\times10^{-9}$  at the pivot scale $k_*=0.05$
Mpc$^{-1}$. A value $\tilde A_s\approx2.4\times 10^{-9}$, as required by Eq.~\rf{Eq:Ratio}, is about $4\sigma$ away from the central value. Albeit not excluded,
this seems statistically problematic. However, the pivot scale assumed in Eq.~\rf{Eq:Ratio} is the horizon scale today, corresponding to $k_*=2.2\times 10^{-4}$
Mpc$^{-1}$, a scale $\sim 200$ times larger than the pivot assumed in the CMB analysis! At such large scales the primordial power spectrum is barely constrained
by CMB or other cosmological probes\footnote{See, e.g., Fig.~6 in Ref.~\refcite{Bringmann:2011}.}, therefore a modification of the order of $10\%$ should be well
within experimental bounds.

One has to keep in mind however that the conventional (geometric) estimation of the acoustic peak position can be influenced by the presence of giant voids, and
one of the CMB anomalies, the Cold Spot, could be indicative of a great void~\cite{Gurzadyan,Szapudi}. This could further relax constrains and make our model even
more plausible.

\section{Conclusion}

In this paper we demonstrate explicitly that in the framework of the $\Lambda$CDM model supplemented in the spherical space with an additional perfect fluid
(namely, quintessence with the constant parameter $w = -1/3$ in the linear equation of state) there is an elegant solution of the horizon problem without
inflation: under the proper choice of the parameters light travels between the antipodal points during the age of the Universe. Consequently, one may suppose that
the observed CMB radiation originates from a very limited space region, which explains its uniformity. Then there seems no need for various inflation scenarios.
In addition this removes any constraints on the uniformity of the universe at the early stage and opens a possibility that the universe was not uniform and that
creation of galaxies and large structures is due to the inhomogeneities that originated in the Big Bang. Besides, in the constructed model the gravitational
potential of any single mass is convergent at any point except for the point of its location, and agreement with the supernovae data is reached. There are certain
serious difficulties when one tries to adjust the proposed concept to the CMB anisotropy arriving at the necessity to change the amplitude of the initial power
spectrum. However, the changes that should be done are well inside experimentally allowed constrains.

\section*{Acknowledgements}

This work is supported by NSF CREST award HRD-1345219 and NASA grant NNX09AV07A. The authors thank Yu.~Shtanov for useful discussions and valuable comments.



\end{document}